\documentstyle[12pt]{article}
\begin{document}
\begin{center}{\Large {\bf Critical Slowing Down along the Dynamic Phase Boundary
in Ising meanfield dynamics}}\end{center}

\vskip 0.5cm

\begin{center}{\it Muktish Acharyya}\\

{Department of Physics, Presidency College}\\

{86/1 College Street, Calcutta-700073, India}\\

{\it {E-mail: muktish.acharyya@gmail.com}}\\

{and}\\

{\it Ajanta Bhowal Acharyya}\\

{Department of Physics, Lady Brabourne College}\\

{P-1/2 Surawardy Avenue, Calcutta-700017, India}\\

{\it {E-mail: ajanta.bhowal@gmail.com}}\end{center}

\vskip 1 cm

\noindent Abstract: 
We studied the dynamical phase transition in kinetic Ising
ferromagnets driven by oscillating magnetic field in meanfield approximation. The meanfield
differential equation was solved by sixth order Runge-Kutta-Felberg method. 
The time averaged magnetisation plays the role of the dynamic order parameter. We studied
the relaxation behaviour 
of the dynamic order parameter close to the transition temperature,
which depends on the amplitude of the applied magnetic field. We 
observed the critical slowing down along the dynamic phase boundary.
We proposed a power law divergence of the relaxation time and estimated the exponent.
We also found its dependence on the field amplitude and compared the result with 
the exact value in limiting case.

\vskip 0.5cm

\noindent {\bf Keywords: Ising model, Meanfield theory, Dynamic transition, 
Relaxation time, Critical slowing down}

\newpage

\noindent {\bf I. Introduction:}

The kinetic Ising model driven by an oscillating magnetic field yields various
nonequilibrium response \cite{rev}.
One interesting nonequilibrium response is the dynamic phase transition. This dynamic phase transition
is widely studied in model ferromagnetic system in the presence of oscillating magnetic field \cite{rev}.
Tome and Oliveira \cite{tom} first observed a prototype of nonequlibrium dynamic transition in the 
numerical solution of meanfield equation of motion for the classical Ising ferromagnet in the presecce of
a magnetic field varying sinusoidally in time. The time averaged (over the complete cycle of the oscillating
magnetic field) magnetisation plays the role of the dynamic order parameter. They \cite{tom} found that this
dynamic ordering depends on the amplitude of the oscillating magnetic field and the temperature of the system.
Systems get dynamically ordered for small values of the temperature and the amplitude of the field. They \cite{tom}
have drawn a phase boundary (separating the ordered and disordered phase) in the temperature field amplitude
plane. More interestingly, they have also reported \cite{tom} a tricritical point on the phase boundary, which
separates the nature (continuous/discontinuous) of the dynamic transition across the phase boundary. This tricritical
point was found just by checking the nature of the transition at all points across the phase boundary. The point where
the nature of transition changes was marked as the tricritical point. No other significance of this tricritical 
point was reported. 
The frequency dependence of this phase boundary was not reported earlier for the dynamic transition in Ising
meanfield dynamics.  

We have studied \cite{bho} numerically the frequency dependence of the
dynamic phase boundary in Ising meanfield dynamics. 
We studied the tricritical behaviour
and found a method of finding the position the 
tricritical point on the dynamic phase boundary. 
The frequency dependence of the position of the 
tricritical point was studied here. We also studied the static
(zero frequency) limit of dynamic phase boundary.

The divergence of the time scale, near the dynamic transition temperature was studied
\cite{mapre1} by Monte Carlo simulation and by solving meanfield dynamical equation
in driven Ising ferromagnet. The critical slowing down was observed in both cases.
In meanfield case, the exponent was also calculated \cite{mapre1} in the limit of
vanishingly small field amplitude. However, it would be interesting to know how this
exponent depends on the field amplitude along the dynamic phase boundary. In this paper,
we have studied this and also checked the earlier result of the value of this exponent
in the limit of vanishingly small field amplitude.

The paper is organised as follows: In the next 
section the model and the method of numerical solution is discussed.
Section III contains the numerical results and 
the paper end with summary of the work in section IV.

\vskip 1 cm

\noindent {\bf II. Model and numerical solution:}

The time ($t$) variation of average magnetisation $m$ of Ising ferromagnet in the
presence of a time varying field, in meanfield approximation, is given as \cite{tom}

\begin{equation}
\tau {{dm} \over {dt}} = -m + {\rm tanh}({{m+h(t)} \over {T}}),
\end{equation}

\noindent where, $h(t)$ is the externally applied sinusoidally oscillating
magnetic field ($h(t) = h_0 {\rm sin}(\omega t)$) and $T$ is the temperature measured
in units of the Boltzmann constant ($K_B$). This equation describes the nonequilibrium
behaviour of instantaneous value of magnetisation $m(t)$ of Ising ferromagnet in 
meanfield approximation. Here, $\tau$ stands for the microscopic relaxation time for
the spin flip \cite{tom}. 
 
In this context, the Hamiltonian describing the Ising ferromagnet may be written
as follows:

\begin{equation}
H=-J\Sigma_{<ij>}S_iS_j-h(t)\Sigma_iS_i
\end{equation}

\noindent where the symbols express their usual meaning \cite{mapre1}.

We have solved this
equation by sixth order Runge-Kutta-Felberg (RKF) \cite{numeric} method to get the instantaneous value
of magnetisation $m(t)$ at any finite temperature $T$, $h_0$ and $\omega (=2\pi f)$.
This method of solving ordinary differential equation ${{dm} \over {dt}} = F(t,m(t))$, is
described briefly as:
\begin{flushleft}{$m(t+dt) = m(t) + \left({{16k_1} \over {135}}+{{6656k_3} \over {12825}}
+{{28561k_4} \over {56430}}-{{9k_5} \over {50}}
+{{2k_6} \over {55}}\right)$}\\
{\rm where}\\
{$k_1 = dt \cdot F(t,m(t))$}\\
{$k_2 = dt \cdot F(t+{{dt} \over 4}, m+{{k_1} \over 4})$}\\
{$k_3 = dt \cdot F(t+{{3dt} \over 8}, m+{{3k_1} \over {32}}+{{9k_2} \over 32})$}\\
{$k_4 = dt \cdot F(t+{{12dt} \over {13}}, m+{{1932k_1} \over {2197}}
-{{7200k_2} \over {2197}}+{{7296k_3} \over 2197})$}
{$k_5 = dt \cdot F(t+dt,m+{{439k_1} \over 216}-8k_2+{{3680k_3} \over 513}
-{{845k_4} \over 4104})$}\\
{$k_6 = dt \cdot F(t+{{dt} \over 2}, m-{{8k_1} \over 27}+2k_2-{{3544k_3} \over 2565}
+{{1859k_4} \over 4104}-{{11k_5} \over 40})$.................................(2)}
\end{flushleft}

\noindent The time interval $dt$ was measured in units of 
$\tau$ (the time taken to flip a single
spin). Actually, we have used $dt = 0.01$ (setting $\tau$=1.0). 
The local error involved in the sixth order RKF method
is of the order of $(dt)^6 (=10^{-12})$. We started with initial condition $m(t=0) = 1.0$. 
In the present case, we kept the frequency $f=0.1$, fixed throughout the study.

\vskip 1cm

\noindent {\bf III. Results:} 

The time averaged magnetisation over a full cycle of the oscillating magnetic field
acts as the role of dynamic order parameter $Q (= {{\omega} \over {2\pi}}
\oint m(t) dt)$. For steady state calculations \cite{bho}, 
this was considered after discarding the 
values of $Q$ for few initial (transient \cite{mapre1}) cycles of the
oscillating field. However, in this paper, we are interested in the transient
behaviour of the dynamic order parameter $Q$. This $Q$ was calculated for every
cycle (of the oscillating magnetic field) starting from the first cycle. The
dynamic order parameter $Q$ was studied as a function of the number of cycles
$n$. This shows a relaxation behaviour of $Q$ (Fig-1). Figure-1 shows the variations
of $log(Q)$ with respect to $n$ for different temperatures. Here, the frequency $f=0.1$
and the field amplitude $h_0=0.2$ remains same. Since, we remain in the disordered
phase the dynamic order parameter $Q$ will vanish as $n$ increases. Here, as the 
temperature $T$ approaches the dynamic transition temperature the relaxation becomes
slower. This is a clear indication of {\it critical slowing down}. Moreover, since
the semilog plot of $Q$ is linear, the relaxation is {\it exponential} and one may
expect the behaviour like $Q \sim exp(-n/{\Gamma})$, where $\Gamma$ defines the
relaxation time. We calculated the relaxation time $\Gamma$ from the least square
fit of the $log(Q)-n$ data (discarding the initial nonlinear part for small $n$). 

For a fixed value of the field amplitude $h_0$, the relaxation time $\Gamma$ was studied
as a function of temperature $T$. It is observed that $\Gamma$ diverges as the temperature
approaches the dynamic transition temperature $T_d(h_0)$. This is demonstrated in 
Figure-2, for three different values of field amplitudes ($h_0$). Here, we observed the
critical slowing down along the dynamic phase boundary. We assumed the scaling law
(if valid still in nonequilibrium case), $\Gamma \sim (T-T_d(h_0))^{-z}$ and estimated
the exponent $z$ as well as $T_d(h_0)$ numerically, for different values of $h_0$.
If our scaling assumption is correct, the variation of
$({{1} \over {\Gamma}})^{({{1} \over {z}})}$ 
with $T$ will be a straight line. The
straight line cuts the temperature axis at $T = T_d(h_0)$. We employed the numerical
method to estimate $z$ and $T_d(h_0)$. The method is as follows: we first use a trial
value of $z$ and calculated
$({{1} \over {\Gamma}})^{({{1} \over {z}})}$ 
for various values of $T$. The data were fitted to a straight line
of the form $y=mx+c$. The error $e = \sum_i (y_i - mx_i-c)^2$ was calculated for
various values of $z$. The accepted value of $z$ was found for which the error $e$
becomes minimum. This is depicted in Figure-3a. Here, $h_0=0.2$ and $e$ becomes minimum
for $z \cong 1$. The best (least square) fit straight line was shown in Figure-3b.
By extrapolating the line one gets $T_d(h_0=0.2)=0.9239$. Similar methods were employed
to get the values of $z$ for other different values of $h_0 = 0.3$ and $h_0=0.4$.
For $h_0=0.3$, we estimated $z=0.993$ and $T_d(h_0=0.3)=0.8115$. 
The corresponding best fit straight line plot is shown
in Figure-4. Figure-5 shows the best fit straight line plot of
$({{1} \over {\Gamma}})^{({{1} \over {z}})}$ 
versus $T$, which estimates $z=0.981$ and $T_d(h_0=0.4)=0.6318$.

\vskip 0.5cm

\noindent {\bf IV. Summary:}

In this paper, we have reported our numerical results of the study of the 
relaxation behaviour of the dynamic order parameter in the dynamic
phase transition in Ising meanfield dynamics. The relaxation was observed to be
exponential in nature and the relaxation time was found to vary as $(T-T_d(h_0))^{-z}$.
We estimated $z$ (as well as $T_d(h_0)$) numerically and found it to vary with
$h_0$. We checked the value of $z$ in the limit of $h_0 \to 0$ obtained
analytically \cite{mapre1}. 

Our present study has the two importances. Firstly, it proves that the critical
slowing down of the dynamic order parameter along the dynamic phase boundary has
power law scaling behaviour (at least in the present case), usually
observed in equilibrium critical phenomena. Secondly, the exponent $z$ is a 
monotonically decreasing function of $h_0$ and 
approaches the exact value ($z=1$) \cite{mapre1}
in the limit $h_0 \to 0$.

\newpage

\setlength{\unitlength}{0.240900pt}
\ifx\plotpoint\undefined\newsavebox{\plotpoint}\fi
\sbox{\plotpoint}{\rule[-0.200pt]{0.400pt}{0.400pt}}%


\vskip 2cm

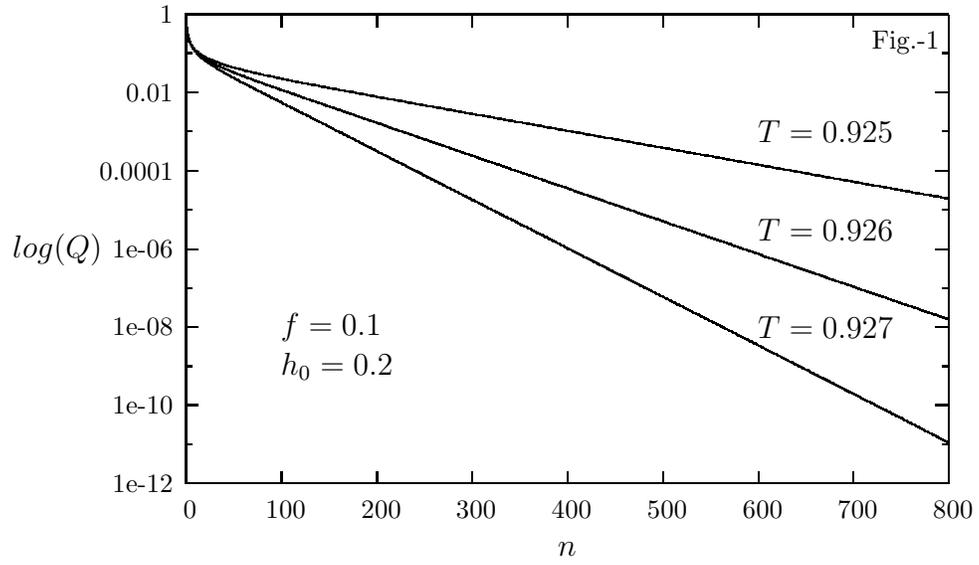
\noindent Fig.-1. The Exponential relaxation of the dynamic order parameter $Q$
at different temperatures. Relaxation is faster at higher temperature.

\newpage

\setlength{\unitlength}{0.240900pt}
\ifx\plotpoint\undefined\newsavebox{\plotpoint}\fi
\sbox{\plotpoint}{\rule[-0.200pt]{0.400pt}{0.400pt}}%
%

\vskip 2cm

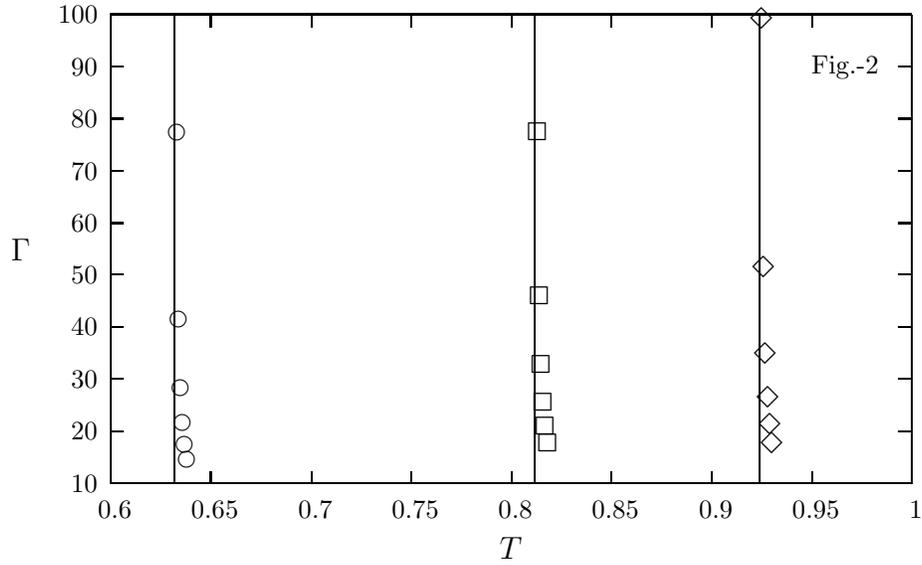
\noindent Fig.-2. The critical slowing down. The relaxation time diverges near the
dynamic transition points $T_d(h_0)$. Different symbols represents different values
of $h_0$. $h_0=0.4(O)$, $h_0=0.3(\Box)$ and $h_0=0.2(\Diamond)$.

\newpage
\setlength{\unitlength}{0.240900pt}
\ifx\plotpoint\undefined\newsavebox{\plotpoint}\fi
\sbox{\plotpoint}{\rule[-0.200pt]{0.400pt}{0.400pt}}%
%

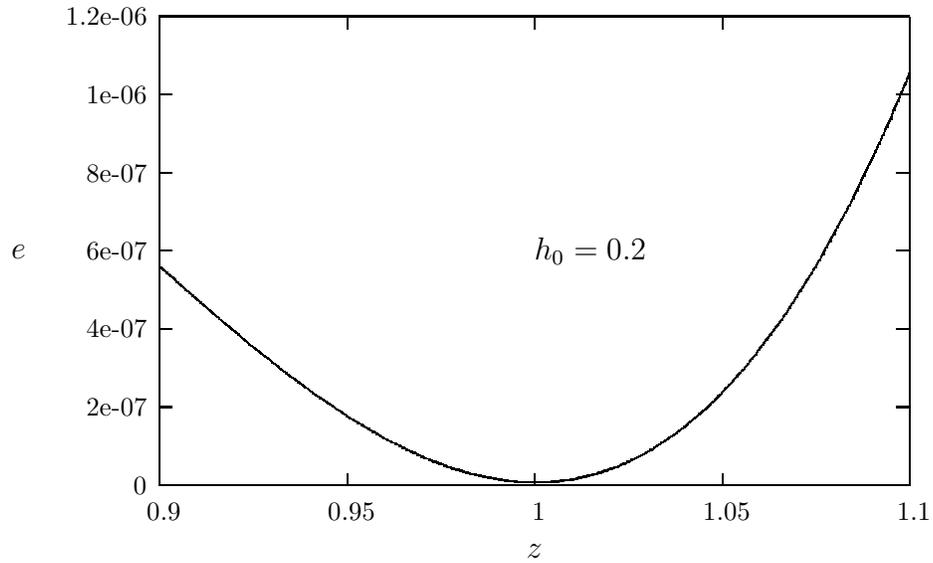
\noindent {Fig.3a.} The variation of the error ($e$) of straight line fit
of $({{1} \over {\Gamma}})^{({{1} \over {z}})}$ versus $T$ for various values of $z$.
Here, $h_0=0.2$. Note that the error is minimum for $z \cong 1.0$. 

\newpage

\setlength{\unitlength}{0.240900pt}
\ifx\plotpoint\undefined\newsavebox{\plotpoint}\fi
\sbox{\plotpoint}{\rule[-0.200pt]{0.400pt}{0.400pt}}%
%

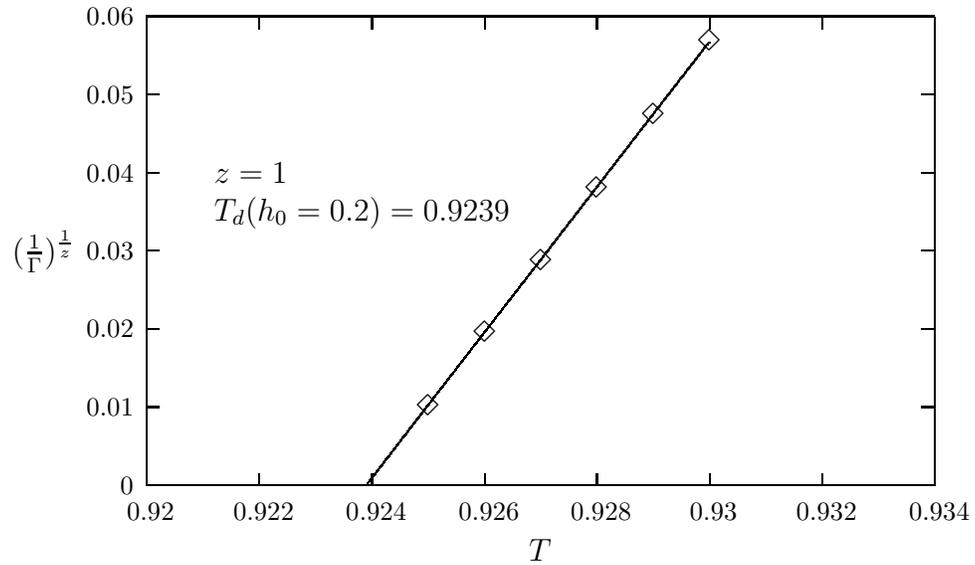
\noindent {Fig.3b.} The best (least error) fit straight line to estimate
$z$ and $T_d(h_0)$ for $h_0=0.2$.

\newpage

\setlength{\unitlength}{0.240900pt}
\ifx\plotpoint\undefined\newsavebox{\plotpoint}\fi
\sbox{\plotpoint}{\rule[-0.200pt]{0.400pt}{0.400pt}}%
%

\noindent {Fig.4.} The best (least error) fit straight line to estimate
$z$ and $T_d(h_0)$ for $h_0=0.3$.

\newpage

\setlength{\unitlength}{0.240900pt}
\ifx\plotpoint\undefined\newsavebox{\plotpoint}\fi
\sbox{\plotpoint}{\rule[-0.200pt]{0.400pt}{0.400pt}}%
%

\noindent {Fig.5.} The best (least error) fit straight line to estimate
$z$ and $T_d(h_0)$ for $h_0=0.4$.

\end{document}